\documentstyle[12pt]{article}
\input epsf

\begin{document}
\renewcommand{\thefootnote}{\fnsymbol{footnote}}
\begin{titlepage}
\renewcommand{\thefootnote}{\fnsymbol{footnote}}
\makebox[2cm]{}\\[-1in]
\begin{flushright}
\begin{tabular}{l}
TUM/T39-97-31 \\
DFTT 73/97
\end{tabular}
\end{flushright}
\vskip0.4cm
\begin{center}
  {\Large\bf
    NLO Corrections to Deeply-Virtual Compton Scattering 
    \footnote{Work supported in
    part by BMBF}}\\ 

\vspace{2cm}

L.\ Mankiewicz\footnote{On leave of absence from N. Copernicus
Astronomical Center, Polish Academy of Science, ul. Bartycka 18,
PL--00-716 Warsaw (Poland)}$^a$, G. Piller$^a$, E. Stein$^b$,
M. V\"anttinen$^a$ and T. Weigl$^a$ 

\vspace{1 cm}

$^a${\em  Physik Department, Technische Universit\"{a}t M\"{u}nchen, \\
D-85747 Garching, Germany}

\vspace{1 cm}

$^b${\em  INFN Sezione di Torino, 
Via P.~Giuria 1, I-10125 Torino, Italy} \\

\vspace{1cm}

{\em \today}

\vspace{1cm}

{\bf Abstract:\\[5pt]} \parbox[t]{\textwidth}{We have calculated the NLO
  corrections to the twist-2 part of the deeply-virtual Compton scattering
  amplitude. Our results for the transverse and antisymmetric parts
  agree with recent calculations by Ji and Osborne and by Belitsky and
  M\"uller. In addition we present NLO results for the  
  longitudinal part of the amplitude.}

\end{center}
\end{titlepage}

\newpage

Our knowledge about the microscopic structure of nucleons is based 
to a large extent  on results from deep-inelastic scattering experiments, 
which study leading twist parton distribution functions \cite{StatusSF}. 
According to factorization theorems \cite{Col89}
these can be linked to forward matrix elements of twist-2 operators.
Forward parton distributions measure the
nucleon response to a process where one parton is removed and subsequently
inserted back into the target along a light-like distance, without changing 
its longitudinal momentum.  
Recently it has been realized that in hard processes where the target 
nucleon recoils elastically, like in deeply virtual Compton
scattering (DVCS) \cite{Ji96,Rad97} 
and hard exclusive lepto-production of mesons 
\cite{Rad97,CFS97,MPW97}, one can study generalized, nonforward nucleon parton
distributions. 
They describe a situation where the removed parton changes its
longitudinal momentum before returning to the nucleon.  
As in the forward case, it is possible to derive exact relations  
between these distributions and matrix elements of QCD operators 
\cite{Ji96,Rad97,Mue97,JiOsb}. 
Some of these matrix elements provide an insight into a gauge-invariant 
decomposition of the nucleon
spin between its constituents \cite{Ji96}. In general it has been recognized
that nonforward parton distributions provide fundamental information about the
nucleon structure and therefore theoretical
\cite{Ji96,Rad97,Mue97,BelMue,Models} and experimental \cite{DVCSExp,Diehl96}
studies of their properties will constitute an important part of future strong
interaction physics.

In the present paper we calculate  NLO $\alpha_s$ corrections to
deeply-virtual Compton scattering amplitude in the ${\overline {\rm MS}}$
scheme. 
For the unpolarized case such a calculation has been 
performed first in Ref.\cite{JiOsb}.
Besides their consequences for future experimental studies of
deeply-virtual Compton scattering, NLO  corrections 
are also of theoretical significance.
Based on the conformal invariance of one-loop QCD 
they can be directly determined from the corresponding corrections 
to the forward scattering amplitude as observed in Ref.\cite{Mue97}.
Indeed, in a recent remarkable paper \cite{BelMue} the full NLO corrections 
have been derived using  conformal symmetry arguments. 
It is gratifying that our results confirm calculations presented 
in \cite{JiOsb,BelMue}. In addition, we present 
the NLO correction to the longitudinal part of the amplitude.

In the following we outline a calculation of the deeply-virtual Compton
scattering amplitude
\begin{equation}
\Pi_{\mu\nu} = i \int\, d^4z\, e^{-i {\bar q} \cdot z} \, 
\left \langle P^\prime \right | T \left[
J_\mu(-{z}/{2}) \,J_\nu ({z}/{2}) \right]\left | P \right \rangle
\label{eq:dvcs_def}
\end{equation}
up to order $\alpha_s$  in asymmetric kinematics
\cite{Wat82,Ditt88,Chen97} with $q (P)$ and $q^\prime (P^\prime)$ being 
the initial and final photon (nucleon) momenta, respectively. 
Within twist-2 accuracy  we set $P^2 = P^{\prime 2} = 0$. 
The virtualities of the initial and final photon are given by $-q^2
= Q^2$ and $-q^{\prime 2} = Q^{\prime 2}$. 
The momentum transfer to the target nucleon is denoted by 
$\Delta = P - P^\prime = q^\prime - q$.  
With the symmetric combinations of initial and
final photon and nucleon momenta,  
${\bar q} = \frac{1}{2} (q + q^\prime)$ and 
${\bar P} =  \frac{1}{2} (P + P^\prime)$,  
we obtain:
\begin{equation}
- {\bar q}^2 = {\bar Q}^2 = \frac{1}{2}(Q^2 + Q^{\prime 2}) +
  \frac{\Delta^2}{4}\, .
\end{equation}
In the following we assume that the virtuality $Q^2$ of the initial photon
is large and allow $Q^{\prime 2}$ to vary between 0 and $Q^2$. 
As a consequence, the asymmetry parameter
\begin{equation}
c = \frac{Q^2 - Q^{\prime 2}}{Q^2 + Q^{\prime 2}}
\label{eq:def_c}
\end{equation}
varies between $c=0$ in the symmetric case of deeply-virtual forward Compton
scattering, and $c=1$ when the final photon is on-shell, i.e. for $Q^{\prime 2}
= 0$.  In addition, we consider the case $ - \Delta^2 \ll Q^2$, and thus
neglect the momentum transfer as compared with ${\bar Q}^2$.  In this limit the
symmetric combination of the initial and final target momentum, ${\bar P}$, can
be considered massless, and only the component of $\Delta$ proportional to
$\bar P$ is relevant:
\begin{equation}
\Delta = 2 \xi {\bar P} + \dots. 
\end{equation} 
It is convenient to introduce the symmetric variable 
${\bar \omega} = \frac{2 {\bar P} \cdot {\bar q}}{{\bar Q}^2}$, 
which is related to the conventional Bjorken
variable $\omega = \frac{2 P \cdot q}{Q^2}$ by
\begin{equation}
{\bar \omega} = (1+c) \omega - c = \frac{c}{\xi} \, .
\end{equation}
Because of the linear relation between $\omega$ and ${\bar \omega}$, analytical
properties of the DVCS amplitude as a function of each variable are closely
related. In particular, at fixed $\Delta^2 = 0$, $\Pi_{\mu\nu}$ is a
holomorphic function of ${\bar \omega}$ with cuts running from $- \infty$ to
$-1$ and from $1$ to $+\infty$. The physical DVCS process corresponds 
to ${\bar \omega}$ approaching the right-hand cut from above,
i.e.  ${\bar \omega} > 1$ having a positive, infinitesimal imaginary part.

In the actual calculation  we follow a method discussed at length in
\cite{Russians}. 
We use dimensional regularization with $d = 4 - 2 \epsilon$ 
for both UV and IR divergences which arise in the massless
twist-2 kinematics described above. 
As explained in \cite{Russians}, the only singularities which appear 
in this approach  arise from the renormalization of twist-2 
quark and gluon operators  which define the parton
distribution functions (\ref{eq:ddist_def_oper}).  
After evaluating the amplitude in $d$-dimensions we have determined 
the finite part of $\Pi_{\mu\nu}$ according to the 
${\overline {\rm MS}}$ scheme.

As a consequence of the operator product expansion the 
DVCS amplitude can be written as \cite{JiOsb}:
\begin{eqnarray}
&& \Pi_{\mu\nu} = (-g_{\mu\nu})^T \bigg [ \sum_q e_q^2 \int_{-1}^1 \, \frac{du}{u}
F_q(u,\xi;{\bar Q}^2) \left( C_0^{S,q}(u{\bar \omega},\xi{\bar \omega}) + a_s\,
C_F \, C_1 ^{S,q}(u{\bar \omega},\xi{\bar \omega}) \right)  \nonumber \\
&& + \int_{-1}^1\, \frac{du}{u} G(u,\xi;{\bar Q}^2)
\,  a_s \, C_1^{S,g}(u{\bar \omega},\xi{\bar \omega}) \bigg] \nonumber \\
&& + i \epsilon_{\mu\nu\alpha\beta} \frac{n^\alpha n^{* \beta}}{n \cdot
n^*} \bigg[\sum_q e_q^2 \int_{-1}^1 \, \frac{du}{u}
\Delta F_q(u,\xi;{\bar Q}^2) \left( C_0^{A,q}(u{\bar \omega},\xi{\bar
\omega})  + a_s \, C_F \, C_1 ^{A,q}(u{\bar \omega},\xi{\bar \omega}) \right) 
\nonumber \\
&& +  \int_{-1}^1\, \frac{du}{u} \Delta G(u,\xi;{\bar Q}^2)\,  
a_s \, C_1^{A,g} (u{\bar \omega},\xi{\bar \omega}) \bigg] \nonumber \\
&& +  \left( {\bar P}_\nu - q_\nu \frac{{\bar P}\cdot {\bar q}}{q^2} \right)
\left( {\bar P}_\mu - q_\mu^\prime \frac{{\bar P}\cdot {\bar q}}{q^{\prime 2}}
\right) \frac{{\bar Q}^2}{2 ({\bar P} \cdot {\bar q})^2} \nonumber \\
&& (1 - {\bar \omega}^2 \xi^2) \left[ \sum_q e_q^2 \int_{-1}^1 \, \frac{du}{u}
F_q(u,\xi;{\bar Q}^2) \, a_s \, C_F \, C_1^{L,q}(u{\bar \omega},\xi{\bar
\omega}) +  
\int_{-1}^1\, \frac{du}{u} G(u,\xi;{\bar Q}^2)\,
a_s \, C_1^{L,g}(u{\bar \omega},\xi{\bar \omega}) \right]  ,\nonumber \\
\label{eq:Pi_pert_exp}
\end{eqnarray}
with $a_s = \frac{\alpha_s({\bar Q}^2)}{2 \pi}$.
The transverse metrics tensor 
$$
(- g_{\mu\nu})^T = \frac{n_\mu n^*_\nu + n_\nu n^*_\mu}{n \cdot n^*} -
g_{\mu\nu} 
$$
and the antisymmetric tensor\footnote{We use the convention of \cite{ItzZub}
for the epsilon tensor.} $\epsilon_{\mu\nu\alpha\beta} 
\frac{n^\alpha n^{*\beta}}{n \cdot n^*}$ are expressed 
through two light-like vectors $n$ and
$n^*$ which fulfill $n^2 = n^{* 2} = 0$, ${\bar P} = \frac{1}{n \cdot n^*}
({\bar P} \cdot n) \, n^*$. 
The nonforward parton distributions
$F(u,\xi;\mu^2), \Delta F(u,\xi;\mu^2), G(u,\xi;\mu^2)$ and $\Delta
F(u,\xi;\mu^2)$ are given by the Fourier transforms of nonforward matrix
elements of corresponding twist-2 QCD string operators 
\cite{Ji96,Rad97,BalBra} normalized at a scale $\mu^2$:
\begin{eqnarray}
F_q(u,\xi;\mu^2) & = & \frac{1}{2} \int \frac{d \lambda}{2 \pi} 
e^{i {\bar P} \cdot n \lambda u}
\left \langle P^\prime \right| {\bar q}(-\frac{\lambda n}{2})
[-\frac{\lambda n}{2};\frac{\lambda n}{2}]
{\hat n} q(\frac{\lambda n}{2}) \left| P \right \rangle_{z^2=0}, \nonumber \\
\Delta F_q(u,\xi;\mu^2) & = & \frac{1}{2} \int \frac{d \lambda}{2 \pi} 
e^{i {\bar P} \cdot n \lambda u}
\left \langle P^\prime \right| {\bar q}(-\frac{\lambda n}{2})
[-\frac{\lambda n}{2};\frac{\lambda n}{2}]
\gamma_5 {\hat n} q(\frac{\lambda n}{2}) \left| P \right \rangle_{z^2=0} ,
\nonumber \\
G(u,\xi;\mu^2) & = &  \frac{1}{u {\bar P}\cdot n}
\int \frac{d \lambda}{2 \pi} 
e^{i {\bar P} \cdot n \lambda u}\, n_\mu n_\nu \,
\left \langle P^\prime \right|
G^{\mu\alpha}(-\frac{\lambda n}{2})[-\frac{\lambda n}{2};\frac{\lambda
n}{2}]G_\alpha^{~\nu} 
(\frac{\lambda n}{2})\left| P \right \rangle_{z^2=0}, \nonumber \\
\Delta G(u,\xi;\mu^2) & = & \frac{1}{u {\bar P}\cdot n}
\int \frac{d \lambda}{2 \pi} 
e^{i {\bar P} \cdot n \lambda u}\, n_\mu n_\nu \,
\left \langle P^\prime \right|
G^{\mu\alpha}(-\frac{\lambda n}{2})[-\frac{\lambda n}{2};\frac{\lambda n}{2}]
{\tilde G}_\alpha^{~\nu} (\frac{\lambda n}{2})\left| P \right \rangle_{z^2=0},
\nonumber \\
\label{eq:ddist_def_oper}
\end{eqnarray}  
with ${\hat n} = n_\mu \gamma^\mu$, and 
$[-\frac{\lambda n}{2};\frac{\lambda n}{2}]$ 
being the path-ordered exponential which ensures gauge-invariance 
nonforward distributions.

The easiest way to present our results is to quote them in the unphysical
domain $\left| {\bar \omega} \right | < 1$ where the amplitude
(\ref{eq:dvcs_def}) is free from imaginary parts arising from the propagation 
of on-shell states. 
There, the required coefficient functions in the ${\overline {\rm MS}}$ 
scheme read
\begin{eqnarray}
C_0^{S,q}(u{\bar \omega},\xi{\bar \omega}) & = & 
\frac{u {\bar \omega}}{1 - u {\bar \omega}} - 
\frac{u {\bar \omega}}{1 + u {\bar \omega}}, \nonumber \\
C_0^{A,q}(u{\bar \omega},\xi{\bar \omega}) & = & 
\frac{u {\bar \omega}}{1 - u {\bar \omega}} + 
\frac{u {\bar \omega}}{1 + u {\bar \omega}}, \nonumber \\
C_1^{S,q}(u{\bar \omega},\xi{\bar \omega}) & = &
         - \frac{9}{2} \, \frac{u {\bar \omega}}{1 - u {\bar \omega}}
         + {{3\,u\,\left( -2\,u + {\it {\bar \omega}}\,{u^2} 
         + {\it {\bar \omega}}\,{{\xi }^2} 
         \right)} \over {2\,\left( 1 - {\it {\bar \omega}}\,u \right) \,
         \left( {u^2} - {{\xi }^2} \right) }} \log(1-{\bar \omega} u)
         \nonumber \\
&& +    
         {{u\,\left( 1 + {{{\it {\bar \omega}}}^2}\,{u^2} - 
         2\,{{{\it {\bar \omega}}}^2}\,{{\xi }^2}
         \right) }\over 
         {2\,{\it {\bar \omega}}\,\left( 1 - {\it {\bar \omega}}\,u \right) \,
         \left( {u^2} - {{\xi }^2} \right) }} 
         \log^2(1-{\bar \omega} u) \nonumber \\
&& +
         {{3\,{u^2}\,\left( 1 - {{{\it {\bar \omega}}}^2}\,{{\xi }^2} 
         \right) }\over {\left( 1 - {{{\it {\bar \omega}}}^2}\,{u^2} 
         \right) \, \left( {u^2} - {{\xi }^2} \right) }}
         \log(1-{\bar \omega} \xi) \nonumber \\
&& +     {{{u^2}\,\left( 1 - {\it {\bar \omega}}\,\xi  \right) \,
         \left( {{{\it {\bar \omega}}}^2}\,{u^2} - 2\,
         {\it {\bar \omega}}\,\xi  - 
         2\,{{{\it {\bar \omega}}}^2}\,{{\xi }^2} -1 \right) }\over 
         {2\,{\it {\bar \omega}}\,
         \left( 1 - {{{\it {\bar \omega}}}^2}\,{u^2} \right) \,\xi \,
         \left( {u^2} - {{\xi }^2} \right) }} 
         \log^2(1-{\bar \omega} \xi)
         \nonumber \\
&& + \,  ( {\bar \omega} \leftrightarrow - {\bar \omega} ), \nonumber \\
C_1^{S,g}(u{\bar \omega},\xi{\bar \omega}) & = & (\sum_q\, e_q^2)\, 
         \frac{1}{2}\, \frac{u^2}{u^2-\xi^2} \left[
         {{4 - 4\,{\it {\bar \omega}}\,u 
         + {{{\it {\bar \omega}}}^2}\,{u^2} - 
         {{{\it {\bar \omega}}}^2}\,{{\xi }^2}}\over 
         {{{{\it {\bar \omega}}}^2}\,\left( {u^2} - {{\xi }^2} \right) }}
         \log(1-{\bar \omega} u) \right. \nonumber \\
&& -     {{2 - 2\,{\it {\bar \omega}}\,u + {{{\it {\bar \omega}}}^2}\,{u^2} - 
         {{{\it {\bar \omega}}}^2}\,{{\xi }^2}}\over 
         {2\,{{{\it {\bar \omega}}}^2}\,\left( {u^2} - {{\xi }^2} \right) }}
         \log^2(1-{\bar \omega} u) \nonumber \\ 
&& +     {{\left( 1 - {\it {\bar \omega}}\,\xi  \right) \,
         \left( {\it {\bar \omega}}\,{u^2} - 4\,\xi  
         - {\it {\bar \omega}}\,{{\xi }^2} \right) }\over
         {{{{\it {\bar \omega}}}^2}\,\xi \,\left( {u^2} - {{\xi }^2} \right) }}
         \log(1-{\bar \omega} \xi)  \nonumber \\
&& -     \left.  {{\left( 1 - {\it {\bar \omega}}\,\xi  \right) \,
         \left( {\it {\bar \omega}}\,{u^2} 
         - 2\,\xi  - {\it {\bar \omega}}\,{{\xi }^2} \right) }\over
         {2\,{{{\it {\bar \omega}}}^2}\,\xi \,
         \left( {u^2} - {{\xi }^2} \right) }}
         \log^2(1-{\bar \omega} \xi) \right] \nonumber \\
&& + \, ({\bar \omega} \leftrightarrow - {\bar \omega}), \nonumber \\
C_1^{A,q}(u{\bar \omega},\xi{\bar \omega}) & = &
             - \frac{9}{2} \, \frac{u {\bar \omega}}
             {1 - u {\bar \omega}} 
             + {{u\,\left( {{{\it {\bar \omega}}}^2}\,{u^2} + 
             3\,{{{\it {\bar \omega}}}^2}\,{{\xi }^2} 
              - 2\,{\it {\bar \omega}}\,u -2 \right) }\over 
             {2\,{\it {\bar \omega}}\,
             \left( 1 - {\it {\bar \omega}}\,u \right) \,
             \left( {u^2} - {{\xi }^2} \right) }} \log(1-{\bar \omega} u)
             \nonumber \\
&& + 
             {{u\,\left( 1 + {{{\it {\bar \omega}}}^2}\,{u^2} 
             - 2\,{{{\it {\bar \omega}}}^2}\,{{\xi }^2}
             \right) }\over {2\,{\it {\bar \omega}}\,
             \left( 1 - {\it {\bar \omega}}\,u \right) \,
             \left( {u^2} - {{\xi }^2} \right) }}  
             \log^2(1-{\bar \omega} u) \nonumber \\
&& +         {{u\,\left( 1 - {\it {\bar \omega}}\,\xi  \right) \,
             \left( 1 + 2\,{{{\it {\bar \omega}}}^2}\,{u^2} 
             + 3\,{\it {\bar \omega}}\,\xi  \right) }\over
             {{\it {\bar \omega}}\,\left( 1 
             - {{{\it {\bar \omega}}}^2}\,{u^2} \right) \,
             \left( {u^2} - {{\xi }^2} \right) }} \log(1-{\bar \omega} \xi)
             \nonumber \\
&& -
             {{u\,\left( 1 - {\it {\bar \omega}}\,\xi  \right) \,
             \left( 1 + {{{\it {\bar \omega}}}^2}\,{u^2} 
             + 2\,{\it {\bar \omega}}\,\xi  \right) }\over 
             {2\,{\it {\bar \omega}}\,\left( 1 
             - {{{\it {\bar \omega}}}^2}\,{u^2} \right) \,
             \left( {u^2} - {{\xi }^2} \right) }}
             \log^2(1-{\bar \omega} \xi) \nonumber \\ 
&& -  \,     ( {\bar \omega} \leftrightarrow - {\bar \omega} ) \, ,\nonumber \\
C_1^{A,g}(u{\bar \omega},\xi{\bar \omega}) & = & (\sum_q\, e_q^2)\, 
             \frac{1}{2}\,\frac{u^2}{u^2-\xi^2} \left[
             {{ \left(  3\,{\it {\bar \omega}}\,{u^2} 
             + {\it {\bar \omega}}\,{{\xi }^2} - 4 \, u \right)}\over
             {{\it {\bar \omega}}\,\left( {u^2} - {{\xi }^2} \right) }}
             \log(1-{\bar \omega} u) \right.
             \nonumber \\
&& + 
             {{ 2\,u - {\it {\bar \omega}}\,{u^2} 
             - {\it {\bar \omega}}\,{{\xi }^2}}\over 
             {2\,{\it {\bar \omega}}\,\left( {u^2} - {{\xi }^2} \right) }}
             \log^2(1-{\bar \omega} u)
             \nonumber \\
&& +         {{4\,u\,\left( 1 - {\it {\bar \omega}}\,\xi  \right) }\over 
             {{\it {\bar \omega}}\,\left( {u^2} - {{\xi }^2} \right) }}
             \log(1-{\bar \omega} \xi)
   -         \left. {{u\,\left( 1 - {\it {\bar \omega}}\,\xi  \right) }\over 
             {{\it {\bar \omega}}\,\left( {u^2} - {{\xi }^2} \right) }}
             \log^2(1-{\bar \omega} \xi) \right]
             \nonumber \\
&& -  \,     ({\bar \omega} \leftrightarrow - {\bar \omega}) \, , \nonumber \\
C_1^{L,q}(u{\bar \omega},\xi{\bar \omega}) & = &
             {{-4\,u}\over {{\it {\bar \omega}}\,
             \left( {u^2} - {{\xi }^2} \right) }}
             \log(1-{\bar \omega} u)
   +         {{4\,{u^2}}\over {{\it {\bar \omega}}\,{u^2}\,\xi  
             - {\it {\bar \omega}}\,{{\xi }^3}}}
             \log(1-{\bar \omega} \xi)
             \nonumber \\
&& + \,      ({\bar \omega} \leftrightarrow - {\bar \omega}) \, , \nonumber \\
C_1^{L,g}(u{\bar \omega},\xi{\bar \omega}) & = & (\sum_q\, e_q^2)\, 
             \frac{1}{2}\, \frac{u^2}{u^2-\xi^2} \left[
             {{8\,\left( 1 - {\it {\bar \omega}}\,u \right) }\over 
             {{{{\it {\bar \omega}}}^2}\,\left( {u^2} - {{\xi }^2} \right) }}
             \log(1-{\bar \omega} u) \right.
             \nonumber \\             
&& + \,      \left. {{4\,\left( {\it {\bar \omega}}\,{u^2} - 2\,\xi  
             + {\it {\bar \omega}}\,{{\xi }^2} \right)}\over
             {{{{\it {\bar \omega}}}^2}\,\xi \,
             \left( {u^2} - {{\xi }^2} \right) }}
             \log(1-{\bar \omega} \xi) \right]
   + \,      ({\bar \omega} \leftrightarrow - {\bar \omega}) \, .
\label{eq:def_coeffs}
\end{eqnarray}
The above expressions for the transverse and antisymmetric coefficient 
functions  agree with the results presented in \cite{JiOsb} and
\cite{BelMue}. 
In the forward limit $\xi = 0$ all coefficients coincide  with the
well-known one-loop corrections to the ${\overline {\rm MS}}$ Wilson
coefficients in deep-inelastic scattering \cite{uDIS,pDIS} 
for the $F_1$, $g_1$ and $F_L$ structure functions. 
As noted already in \cite{JiOsb}, the transverse and
antisymmetric coefficient functions are regular in the limit 
$Q^{\prime 2} \rightarrow 0$, i.e.  ${\bar \omega} \xi \to 1$. 
After multiplying by the longitudinal photon polarization vectors 
$\epsilon_{L\mu}^*(q^\prime)$ and
$\epsilon_{L\nu}(q)$, the complete longitudinal amplitude
\begin{eqnarray}
&&\epsilon_{L\mu}^*(q^\prime) \, \Pi_{\mu\nu} \, \epsilon_{L\nu}(q) =
\frac{1}{2}(1- {\bar \omega}^2 \xi^2)^{1/2} \nonumber \\
&& \left[ \sum_q e_q^2 \int_{-1}^1 \, \frac{du}{u}
F_q(u,\xi;{\bar Q}^2) \, a_s \, C_1^{L,q}(u{\bar \omega},\xi{\bar \omega}) + 
\int_{-1}^1\, \frac{du}{u} G(u,\xi;{\bar Q}^2)\,
a_s \, C_1^{L,g}(u{\bar \omega},\xi{\bar \omega}) \right]  ,\nonumber \\
\label{eq:ampl_long}
\end{eqnarray}
vanishes when one of the photons goes on shell.

A few comments are in order:
to isolate the gluonic coefficients $C_1^{i,g}, \, i = S,A,L$
we found it advantageous to work in light-cone gauge $n\cdot A = 0$. 
Here one can formally express the non-forward matrix element of the 
product of two gluon fields by a combination of non-forward parton 
distributions \cite{Rad97,JiJaffe}:
\begin{eqnarray}
\left \langle P^\prime \right| A^A_\rho(-\frac{z}{2})
A^A_\sigma(\frac{z}{2}) \left| P \right \rangle_{z^2=0} & = & \frac{1}{d-2}
(-g_{\rho\sigma})^T 
\int_{-1}^1\, du\, \frac{u\, G(u,\xi;\mu^2)}{(u-\xi+i\epsilon)
(u+\xi-i\epsilon)} e^{-i u {\bar P}\cdot z} \nonumber \\
& - & \frac{1}{d-2} \, i \epsilon_{\rho\sigma\alpha\beta} \frac{n^\alpha n^{*
\beta}}{n \cdot n^*} 
\int_{-1}^1\, du\, \frac{u\, \Delta G(u,\xi;\mu^2)}{(u-\xi+i\epsilon)
(u+\xi-i\epsilon)} e^{-i u {\bar P}\cdot z} . \nonumber \\
\label{eq:def_glu_dist_d_dim}
\end{eqnarray}
The factor $\frac{1}{d-2}$ in the second line arises from the relation
between the product of the gluon field strength tensor and its dual,  
and the polarized gluon distribution (see e.g. \cite{SPM97}). 
Here a product of two epsilon tensors arises which can be expressed 
by a determinant of metric tensors in $d$ dimensions. 
Although within such a scheme unpolarized and polarized contributions in 
Eq.(\ref{eq:def_glu_dist_d_dim}) are treated symmetrically, 
the factor $\frac{1}{d-2}$ is 
customary set to its $d=4$ value in the current literature
\cite{vanNeerven}. 
As different choices correspond merely to different
factorization schemes\footnote{We are grateful to W. Vogelsang for discussions
about this point.}, the result for $C_1^{A,g}$ presented in
(\ref{eq:def_coeffs}) 
has been calculated according to the standard convention.

In the case of $C_1^{A,q}$ there are well-know complications \cite{Larin1}
related to the 'tHooft-Veltman definition of the $\gamma_5$ matrix in $d$
dimensions \cite{tHV}. 
In order to avoid a non-vanishing two-loop anomalous
dimension of the flavor non-singlet axial current operator we have followed
Ref.\cite{Larin2} and imposed an additional, 
finite renormalization of the twist-2
axial string operator $A(-\frac{z}{2},\frac{z}{2}) = {\bar
  q}(-\frac{z}{2})[-\frac{z}{2};\frac{z}{2}] \gamma_5 {\hat z} q(\frac{z}{2})$.
It can be derived from the condition \cite{Larin1} that the renormalized
vertices of the operators $A(-\frac{z}{2},\frac{z}{2})$ and
$V(-\frac{z}{2},\frac{z}{2})= {\bar q}(-\frac{z}{2})[-\frac{z}{2};\frac{z}{2}]
{\hat z} q(\frac{z}{2})$ coincide up to the $\gamma_5$ matrix:
\begin{equation}
{\tilde Z}_5^{ns} \otimes R_{\overline {\rm MS}} \left \langle {\bar \psi} \, A
\, \psi \right \rangle = \gamma_5 R_{\overline {\rm MS}} \left \langle {\bar
\psi} \, V \, \psi \right \rangle
\label{eq:def_Z_ns}
\end{equation}
Here $R_{\overline {\rm MS}}$ denotes the R-operation,  i.e. 
the subtraction of the UV singularity in the ${\overline {\rm MS}}$ scheme.
The symbol $\otimes$ in the above equation stands for the  convolution of 
the kernel ${\tilde Z}_5^{ns}$ with the operator $A$ according to:
\begin{equation}
{\tilde Z}_5^{ns} \otimes A = \int_0^1 d\lambda \int_0^{\bar \lambda} d \rho 
\, {\tilde Z}_5^{ns} \, A(-\frac{z}{2}(1-2 \lambda),\frac{z}{2}(1-2 \rho))
\end{equation}
with ${\bar \lambda} = 1 - \lambda$,  and 
to one loop accuracy,
\begin{equation}
{\tilde Z}_5^{ns} = \delta({{\bar \lambda}})\, \delta({{\bar \rho}}) + a_s (- 8
C_F) \, .
\label{eq:Z_ns}
\end{equation}
The condition (\ref{eq:def_Z_ns}) is clearly inappropriate in the case of the
flavor-{\em singlet} axial current, which acquires an anomalous dimension at
the two-loop level due to the axial anomaly \cite{Larin2,Anomaly}.
We have therefore undone the effect of (\ref{eq:Z_ns}) in the flavor-singlet
coefficient $C_1^{A,q}$ by subtracting the term proportional to ${\bar \omega}$
in the DVCS amplitude resulting from ${\tilde Z}_5^{ns}$.  Subsequently we have
replaced it by a contribution which follows from the application of the
renormalization constant of the flavor singlet axial current operator.  This
has been determined to two-loop accuracy in Ref.\cite{Larin1} from the
requirement that the anomaly equation is preserved at higher orders.  After
this correction has been incorporated it turns out that the one-loop
coefficients $C_1^{A,q}$ are the same in the flavor singlet and non-singlet
channels.  Different choices of normalizations of the flavor singlet axial
current can again be related to different factorization schemes. A physical
quantity, like the antisymmetric part of the deeply-virtual Compton scattering
amplitude, does not depend on the choice of $\tilde{Z}_5^s$, provided one
consistently defines the NLO evolution equations \footnote{We thank S.~Forte
  for discussions about this point.}.

It is perhaps worthwhile to note that at this point one can fully appreciate
the power of the conformal approach \cite{BelMue}, which allows to determine
the coefficient functions (\ref{eq:def_coeffs}) from the well-known one-loop
Wilson coefficients in the forward case.

We have presented a calculation of NLO corrections to the deeply-virtual 
Compton scattering amplitude. 
An analysis of future experimental data requires an 
analytical continuation of Eq.(\ref{eq:Pi_pert_exp}) to the physical domain 
${\bar \omega} > 1$, and the inclusion of NLO effects in the evolution of
non-forward parton densities. 
The latter are known at present only for the non-singlet
sector \cite{NLOEvol}. 
However, even before a complete NLO analysis is completed, 
it will be interesting to see, using perhaps some phenomenological model for
non-forward parton distributions \cite{Models,MPW97}\footnote{The
  phenomenological parametrization used in ref.\cite{MPW97} has been proposed
  by A.~Radyushkin}, to which extent corrections discussed in this paper
modify the leading-order results.

\vskip 1cm

{\bf Acknowledgments}\\
This work was supported in part by BMBF and by KBN grant
2~P03B~065~10. E.S.~thanks Deutsche Forschungsgemeinschaft (DFG) for 
financial support. We thank Dieter M\"uller for pointing out
an error in Eq.(8).
\vfill\eject

\clearpage


\end{document}